\documentclass[12pt,english]{article}
\usepackage[utf8]{inputenc}
\usepackage[english]{babel}
\usepackage{float,caption,subcaption}
\usepackage{booktabs,longtable}
\usepackage{amsmath}
\usepackage{amssymb,amsthm,amsfonts}
\usepackage{enumerate}
\usepackage{tikz}
\usepackage[authoryear,round]{natbib}
\usepackage[hyphens]{url}
\usepackage{hyperref}
\usepackage[a4paper,body={16cm,23.7cm}]{geometry}

\hypersetup{colorlinks=true, linkcolor=cyan, citecolor=cyan, urlcolor=black, breaklinks=true}
\newtheorem{theorem}{Theorem}
\newtheorem{lemma}{Lemma}

\newtheorem{proposition}{Proposition}

\begin{document}

\title{\textbf{Historical claims problems}\thanks{We thank financial support from MCIN/AEI/10.13039/501100011033/ FEDER, UE, through grants PID2022-137211NB-I00 and PID2023-147391NB-I00 and PID2023-146364NB-I00.
This research was also partially supported by the CIPROM/2024/34 grant, funded by the Conselleria de Educación, Cultura, Universidades y Empleo, Generalitat Valenciana.}}

\author{
\textbf{Juan C. Gon\c calves-Dosantos}\thanks{Universidad Miguel Hern\'{a}ndez.} \\
\textbf{Ricardo Mart\'{\i}nez}\thanks{Universidad de Granada.} \\
\textbf{Juan D. Moreno-Ternero}\thanks{Universidad Pablo de Olavide.} \\
\textbf{Joaqu\'{\i}n S\'{a}nchez-Soriano}\thanks{Universidad Miguel Hern\'{a}ndez.}} 

\maketitle

\begin{abstract}
We explore the resolution of claims problems with history. At a given period of time, a group of agents holds claims over an insufficient endowment, as they did in previous periods. The solution to the present-period problem might be influenced by the solutions at previous-periods problems (history). We introduce a natural \textit{historical operator}, which extends standard rules (solving one-shot claims problems) to construct rules that solve claims problems with history. We study the preservation of properties by this operator and also obtain a characterization result for it. 
\end{abstract}

\noindent \textbf{\textit{JEL numbers}}\textit{: D63.}\medskip{}

\noindent \textbf{\textit{Keywords}}\textit{: sharing, claims, axioms, bankruptcy, operator, history.}\medskip{}  \medskip{}

\newpage

\section{Introduction} 
\cite{ONeill1982} is the seminal paper for the sizable literature on \textit{claims problems} (also known as bankruptcy problems, or rationing problems).\footnote{The reader is referred to \cite{Thomson03, Thomson15, Thomson19} for focal surveys of this literature.} 
His basic and stylized model formalized a group of individuals having (conflicting) claims over an insufficient amount of a perfectly divisible good. More precisely, a claims problem is described by a triplet $(N,c,E)$, where $N$ is a group of agents, $c_{i}\in \mathbb{R}_+$ is the claim of each agent $i\in N$, $c\equiv(c_i)_{i\in N}$ is the resulting profile of claims, $E\in \mathbb{R}_+$ is the endowment that has to be allocated among agents in $N$, and the condition $E\le\sum_{i\in N}c_{i}$ is satisfied. The issue was to determine rules that would associate with each of these problems a specific allocation of the endowment satisfying the conditions of \textit{boundedness} (no agent receives a negative amount, or a larger amount than his claim) and \textit{balance} (the aggregate allocation is equal to the endowment). 

The field of operations research has devoted considerable attention to this model \citep[e.g.,][]{Lahiri1994, Annabi2012, Brink2013, GimenezGomez2014, Brink2017, SaavedraNieves2020}, some of its applications \citep[e.g.,][]{CasasMendez2011, Gutierrez2018, Bergantinos2025, Acosta2021}, or several of its generalizations \citep[e.g.,][]{Calleja2005, Bergantinos2008, Bergantinos2010, Estan2021, EstevezFernandez2021, Martinez22}. 
But little attention has been paid to the extension of the model to a dynamic setting, in which agents face a claims problem at each period of time.\footnote{A recent exception along these lines is \cite{MorenoTernero21}, which we shall discuss further later.} Protocols for the reduction of pollution, food rationing in refugee camps, university budgeting procedures, water distribution in drought periods, or even some resource allocation procedures in the public health care sector \citep[e.g.,][]{Pulido02, Pulido08, Ju21, AcostaVega2023} are examples that would actually fit the framework of dynamic claims problems. 

This paper aims to contribute in that direction by exploring the resolution of \textit{claims problems with history}, or \textit{historical claims problems}. More precisely, in our model there exists a fixed group of agents who hold claims over an insufficient endowment at a given period of time. That is, they face a claims problem in a given period of time. But we assume they were also facing successive claims problems in previous periods (history). It seems natural to assume that the solution to the problem at the present period might be influenced by the solutions of the problems they faced in previous periods. Our main contribution will be to introduce (and characterize) a natural \textit{historical operator} to extend standard rules (solving one-shot claims problems) in order to construct rules that solve claims problems with history. We shall also study how this operator preserves properties. That is, whether the rule induced by the operator inherits some of the properties the original rule satisfied.

The concept of operators for the space of rules in (standard) claims problems was first introduced by \cite{Thomson08}. They considered three operators (the \textit{duality operator}, the \textit{claims truncation operator}, and the \textit{attribution of minimal rights operator}), which assigned a new rule to any rule to solve (standard) claims problems. They established a number of results linking them and determined which properties of rules were preserved under each of these operators, and which were not (as we do in this paper). \cite{Hougaard12} extended this analysis upon studying a family of operators generalizing two in \cite{Thomson08}. The family was inspired by the so-called \textit{composition} properties, principles that are frequently employed in the literature on claims problems, pertaining to the way rules react to tentative allocations of the available amount \citep[e.g.,][]{Young1988, Moulin00}. 

\cite{Hougaard13, Hougaard13b} focused instead on \textit{extension operators}, which associate with each rule to solve standard claims problems a rule to solve \textit{extended} claims problems in which agents also have \textit{baselines} (beyond claims) which might play a role in the allocation process.\footnote{See also \cite{Timoner2016} or \cite{Izquierdo2025} for similar concepts to baselines.} The historical operator we introduce in this paper can also be seen as an extension operator, but in this case to solve claims problems with history, rather than claims problems with baselines.

Finally, \cite{MorenoTernero21} take an intermediate step between the claims problems with history we analyze in this paper and the claims problems with baselines mentioned above. While also facing a dynamic setting for claims problems, \cite{MorenoTernero21} assume that, at each period, the corresponding claims problem is enriched by an index summarizing the amounts each agent obtained in the previous periods. The index could have many forms, ranging from the (arithmetic or geometric) average to some lower or upper bounds, as well as simply the choice of a specific period. As such, it could also be interpreted as a baselines profile. 

In this paper, we take first a {\it direct approach} to analyze historical claims problems.\footnote{The terminology is borrowed from \cite{Thomson19}.}
That is, we single out a natural class of rules to solve those problems. In
short, rules within this class can be described via a three-stage procedure. First, consider the \textit{history-adjusted claims} vector, obtained after adding to the present-day claim all past losses or deviations each agent may have faced at each past period. Second, obtain a tentative allocation by applying a standard rule to the resulting standard problem with those history-adjusted claims. And third, make the necessary adjustments to guarantee that the final allocation satisfies balance and boundedness.

In other words, the class arises after submitting the domain of standard rules to a
{\it historical operator} reflecting the three-stage procedure described above. We study the preservation of focal properties under such an operator, which can also be interpreted as the study of the robustness of the class.

We then take an {\it axiomatic approach} and study the implications
of new axioms reflecting ethical or operational principles in this
general context (of historical claims problems). We provide an axiomatic
characterization for the class of rules just
described. More precisely, we show that the historical operator described above is characterized by the combination of two axioms: \textit{balanced treatment} and \textit{non-arbitrariness}. The former says that if two agents are not fully honored, the marginal contribution the operator yields must be the same for both. The latter says that if the operator fully honors an agent's claim, this does not come at the expense of harming those whose claims are not fully honored. An interesting step of our characterization result is to show that the combination of these two axioms implies another condition, known as \textit{present boundedness}. This says that if the standard rule (applied to the history-adjusted claims) stipulates that an agent should be awarded an amount exceeding his claim, then this agent should be fully honored when considering the historical context.

The rest of the paper is organized as follows. In Section 2, we
describe the basic framework for standard claims problems, as
well as the new one to address historical claims problems. In Section 3, we present our operator and study the preservation of axioms it yields. In Section 4, we derive the operator axiomatically. We conclude in Section 5 with some further insights. 

\section{Model and basic concepts\label{Model}}

\subsection{Claims problems}

We study historical claims problems in a variable-population model. The set of potential claimants, or \textbf{agents}, is identified with the set of natural numbers $\mathbb{N}$. Let $\mathcal{N}$ be the class of finite subsets of $\mathbb{N}$, with generic element $N=\{1,\ldots, n\}$. For each $i\in N$, let $c_{i}\in\mathbb{R}_{+}$ be $i$'s \textbf{claim} and $c\equiv(c_{i})_{i\in N}$ the claims profile.\footnote{For each $N\in\mathcal{N}$, each $M\subseteq N$, and each $z\in\mathbb{R}^{n}$, let $z_{M}\equiv(z_{i})_{i\in M}$. For each $i\in N$, let $z_{-i}\equiv z_{N\setminus\{i\}}$. Also, let $\mathbf{1}_n=(1,1,\dots,1) \in \mathbb{R}^{N}$. 
Finally, for each pair $x,y \in \mathbb{R}^N$, let $\min\{x,y\}=(\min\{x_i,y_i\})_{i=1}^n$.} 
A (claims) \textbf{problem} is a triple $(N,c,E)$ consisting of a population $N\in\mathcal{N}$, a claims profile $c\in\mathbb{R}_{+}^{n}$, and an \textbf{endowment} $E\in\mathbb{R}_{+}$ such that $\sum_{i\in N}c_{i}\geq E$. Let $C\equiv\sum_{i\in N}c_{i}$. To avoid unnecessary complication, we assume $C>0$. Let $\mathcal{D}^{N}$ be the set of rationing problems with population $N$ and $\mathcal{D}\equiv\bigcup_{N\in\mathcal{N}}\mathcal{D}^{N}$.

Given a problem $\left(N,c,E\right)\in\mathcal{D}^{N}$, an \textbf{allocation} is a vector $x\in\mathbb{R}^{N}$ satisfying 
\textit{boundedness} (for each $i\in N$, $0\leq x_{i}\leq c_{i}$) and 
\textit{balance} ($\sum_{i\in N}x_{i}=E$). 
A \textbf{(standard) rule} on $\mathcal{D}$ is a mapping, $R: \mathcal{D} \longrightarrow \bigcup_{N\in\mathcal{N}} \mathbb{R}^{N}$, that associates with each problem $\left(N,c,E\right)\in\mathcal{D}$ an allocation $R\left(N,c,E\right)\in\mathbb{R}_{+}^{N}$. Let $\mathcal{R}$ denote the set of all (standard) rules so constructed. 

Some classical rules are the \textit{proportional} rule, which yields allocations proportionally to claims; the \textit{constrained equal-awards} rule, which distributes the endowment equally among all agents, subject to no agent receiving more than his claim; the \textit{constrained equal-losses} rule, which makes losses as equal as possible, subject to no one receiving a negative amount; and the \textit{Talmud} rule, which is a hybrid between the last two. Formally,

\textbf{Proportional rule}. For each $(N,c,E) \in \mathcal{D}$ and each $i \in N$,
$$
R^{PROP}_i(N,c,E) = \frac{c_i}{C} E.
$$

\textbf{Constrained equal-awards rule}. For each $(N,c,E) \in \mathcal{D}$ and each $i \in N$,
$$
R^{CEA}_i(N,c,E)=\min\{c_i ,\delta\},
$$
where $\delta \in \mathbb{R}$ is such that $\sum_{i \in N} \min\{c_i, \delta\} = E$.

\textbf{Constrained equal-losses rule}. For each $(N,c,E) \in \mathcal{D}$ and each $i \in N$,
$$
R^{CEL}_i(N,c,E)=\max\{0,c_i -\delta\},
$$
where $\delta \in \mathbb{R}$ is such that $\sum_{i \in N} \max\{0,c_i - \delta\} = E$.

\textbf{Talmud rule}. For each $(N,c,E) \in \mathcal{D}$ and each $i \in N$,
$$
R^{TAL}_i(N,c,E)= 
\begin{cases}
\min \left\{\frac{1}{2} c_i, \delta \right\} & \text { if } E \leq \frac{1}{2} C \\ \max \left\{\frac{1}{2} c_i, c_i-\gamma \right\} & \text { if } E \geq \frac{1}{2} C\end{cases}
$$
where $\delta$ and $\gamma$ are such that $\sum_{i \in N} R^{TAL}_i(N,c,E)=E$.

\begin{figure}
\centering
\begin{subfigure}{0.24\textwidth}
\begin{tikzpicture}[scale=0.8]
        \draw[->] (-0.2,0) -- (3.5,0) node[below] {$x_1$};
        \draw[->] (0,-0.2) -- (0,6.5) node[left] {$x_2$};
        \fill (3,6) circle (2pt) node[above right] {$c$};
        \draw[dashed] (0,6) -- (3,6) -- (3,0);
        \draw[orange] (0,0) -- (3,6);
\end{tikzpicture}
\caption{ }
\end{subfigure}
\begin{subfigure}{0.24\textwidth}
\begin{tikzpicture}[scale=0.8]
        \draw[->] (-0.2,0) -- (3.5,0) node[below] {$x_1$};
        \draw[->] (0,-0.2) -- (0,6.5) node[left] {$x_2$};
        \fill (3,6) circle (2pt) node[above right] {$c$};
        \draw[dashed] (0,6) -- (3,6) -- (3,0);
        \draw[orange] (0,0) -- (3,3) -- (3,6);
\end{tikzpicture}
\caption{}
\end{subfigure}
\begin{subfigure}{0.24\textwidth}
\begin{tikzpicture}[scale=0.8]
        \draw[->] (-0.2,0) -- (3.5,0) node[below] {$x_1$};
        \draw[->] (0,-0.2) -- (0,6.5) node[left] {$x_2$};
        \fill (3,6) circle (2pt) node[above right] {$c$};
        \draw[dashed] (0,6) -- (3,6) -- (3,0);
        \draw[orange] (0,0) -- (0,3) -- (3,6);
\end{tikzpicture}
\caption{}
\end{subfigure}
\begin{subfigure}{0.24\textwidth}
\begin{tikzpicture}[scale=0.8]
        \draw[->] (-0.2,0) -- (3.5,0) node[below] {$x_1$};
        \draw[->] (0,-0.2) -- (0,6.5) node[left] {$x_2$};
        \fill (3,6) circle (2pt) node[above right] {$c$};
        \draw[dashed] (0,6) -- (3,6) -- (3,0);
        \draw[orange] (0,0) -- (1.5,1.5) -- (1.5,4.5) -- (3,6);
\end{tikzpicture}
\caption{}
\end{subfigure}
\caption{Path of awards of the (a) proportional, (b) constrained equal awards, (c) constrained equal losses, and (d) Talmud rules. \label{fig_paths_standard}}
\end{figure}
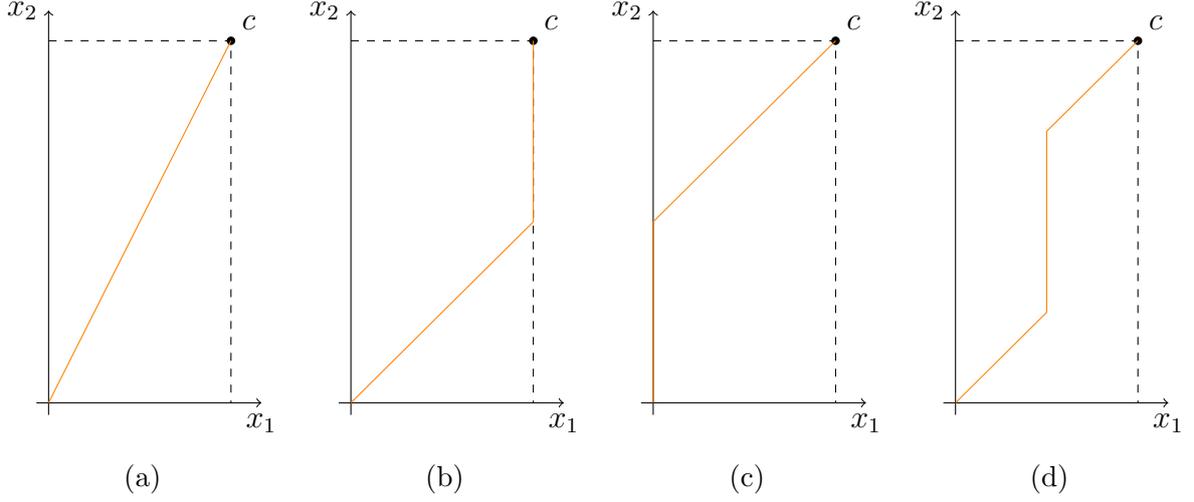

Given a claims vector $c \in \mathbb{R}^N_+$, the locus of the allocation chosen by a rule $R$ as the endowment varies from 0 to $C$ is the \textbf{path of awards} of the rule $R$ for the claims vector $c$, denoted by $p^{R}(c)$. In Figure \ref{fig_paths_standard}, we illustrate the paths of awards for the four most prominent rules mentioned above.

Rules are typically evaluated in terms of the properties (axioms) they satisfy. The literature has provided a wide variety of axioms reflecting ethical or operational principles. We concentrate here on the main ones.\footnote{The reader is referred to \cite{Thomson19} for further axioms and details about them.} 

We start with the basic impartiality requirement of allotting equal amounts to agents with equal claims. Formally, a rule satisfies \textbf{equal treatment of equals} if for each $\left(N,c,E\right)\in\mathcal{D}$, and each pair $i,j\in N,$ we have $R_{i}\left(N,c,E\right)=R_{j}\left(N,c,E\right),$ whenever $c_i = c_j$.
A strengthening says that agents with larger claims receive and lose at least as much as agents with smaller claims. That is, a rule is \emph{order preserving}, if for each $\left(N,c,E\right)\in\mathcal{D}$ and $i,j\in N$, $c_{i}\geq c_{j}$ implies that $R_{i}\left(N,c,E\right)\geq R_{j}\left(N,c,E\right)$. Also, $c_{i}-R_{i}\left(N,c,E\right)\geq c_{j}-R_{j}\left(N,c,E\right)$. The first inequality is referred to as \textbf{order preservation in gains}, and the second as \textbf{order preservation in losses}.
Another strengthening says that a permutation of claimants should yield the same permutation of awards. That is, a rule is \textbf{anonymous} if for all $\left(N,c,E\right)\in\mathcal{D}$, $\pi\in\Pi^{N},$ and $i\in N$, $R_{\pi(i)}\left(N,(c_{\pi(i)})_{i\in N},E\right)=R_{i}\left(N,c,E\right)$.

Some technical properties come next. First, \textbf{scale invariance}, which says that if claims and endowment are multiplied by the same positive number, then so should all awards. That is, for each $(N,c,E)\in\mathcal{D}$ and each $\rho >0,$ $R(N,\rho c,\rho E)=\rho R(N,c,E).$ Second, \textbf{continuity}, which says that if, for each sequence $\left\{\left(N,c^k,E^k\right)\right\}$ of problems in $\mathcal{D}$, and each $\left(N,c,E\right)\in \mathcal{D},$ if $\left(N,c^k,E^k\right)\to\left(N,c,E\right)$, then $R\left(N,c^k,E^k\right)\to R\left(N,c,E\right)$. Third, \textbf{self-duality}, which says that for each $(N,c,E)\in\mathcal{D}$, $R\left(N,c,C-E\right)=c-R\left(N,c,E\right)$.

The next axiom 
states that each agent's award should not be smaller than $\frac{1}{n}$-th of his truncated claim.\footnote{This axiom was introduced by \cite{MorenoTernero04}.}  Formally, a rule satisfies \textbf{securement} if for each $\left(N,c,E\right)\in \mathcal{D}$ and each $i \in N$, $R_i(N,c,E) \ge \frac{1}{n} \min \left\{c_i,E \right\}$.

We now consider a pair of axioms pertaining to wrong estimations of the endowment from above or from below.\footnote{These two axioms were initially studied by \cite{Moulin87} and \cite{Young1988}. See also \cite{Moulin00} and \cite{Chambers2017} for more recent uses.} 
A rule $R$ satisfies \textbf{composition up} if, for each $(N,c,E)\in  \mathcal{D},$ and each pair $E_{1},E_{2}\in \mathbb{R}_{++}$ such that $E_{1}+E_{2}=E,$ $R(N,c,E)=R(N,c,E_{1})+R\left( N,c-R(N,E_{1},c),E_{2}\right)$.
A rule $R$ satisfies \textbf{composition down} if, for each $(N,c,E)\in \mathcal{D},$ and each $E^{\prime }>E,$ we have $R(N,c,E)=R\left( N,R(N,E^{\prime },c),E\right)$.

We conclude with several formulations of the principle of \textit{solidarity}.\footnote{Solidarity properties with respect to population changes were introduced by \cite{Thomson83a,Thomson83b} in related models. \cite{Roemer86} introduced the solidarity notion referring to the available resource. \cite{Chun99} and \cite{MorenoTernero06b} studied solidarity notions combining both aspects. Finally, \cite{Thomson12b} is an excellent reference for the many uses of the notion of consistency and its solidarity underpinnings.} The first one says that when there is more to be divided, nobody should lose. Formally, a rule $R$ is \textbf{resource monotonic} if, for each $\left(N,c,E\right)\in\mathcal{D}$ and each $E^{\prime }>E$, with $E^{\prime }\le\sum c_{i}$, we have $R(N,c,E)\leq R(N,c,E^{\prime })$. The second one says that if an agent's claim increases, she should receive at least as much as she did initially. Formally, a rule $R$ is \textbf{claims monotonic} if, for each $\left(N,c,E\right)\in\mathcal{D}$, each $i\in N$, and each $ c_{i}^{\prime }>c_{i}$, we have $R_{i}(N,(c_{i}^{\prime},c_{N\setminus\{i\}}),E)\geq R_{i}(N,(c_{i},c_{N\setminus\{i\}}),E)$. 
The third one says that if new claimants arrive, each claimant initially present should receive at most as much as she did initially. 
Formally, a rule $R$ is \textbf{population monotonic} if, for each $\left(N,c,E\right)\in\mathcal{D}$ and $\left(N^{\prime},c^{\prime},E\right) \in \mathcal{D}$ such that $N\subseteq N^{\prime}$ and $c_{N}^{\prime}=c$, then $R_{i}\left(N^{\prime},c^{\prime},E\right) \leq R_{i}\left(N,c,E\right)$, for each $i\in N$. 
Finally, the last axiom relates the allocation of a given problem to the allocations of the subproblems that appear when we consider a subgroup of agents as a new population and the amounts gathered in the original problem as the available endowment. 
Formally, a rule $R$ is \textbf{consistent} if for each $\left(N,c,E\right)\in\mathcal{D}$, each $M\subset N,$ and each $i\in M,$ we have $R_{i}\left(N,c,E\right)=R_{i}(M,c_{M},E_{M}),$ where $E_{M}=\sum_{i\in M}R_{i}(N,c,E)$. 

\subsection{Historical claims problems}

We now enrich the model to include a historical context. We assume that there are $T=\{1,\dots,|T|\}$ previous periods of time. In each $t \in T$, agents in $N$ faced a claims problem $(N,c^{(t)},E^{(t)}) \in \mathcal{D}$, which was solved by means of an allocation $x^{(t)} \in \mathbb{R}^N$. This allocation satisfies boundedness and balance, but no information is given about the mechanism that was used to determine it; we only know how the previous claims problems have been solved. A historical context, or \textbf{history}, is a sequence of claims problems and allocations (over time). That is, $h=\left\{ \left( N,c^{(t)},E^{(t)},x^{(t)} \right) \right\}_{t \in T}$. For ease of notation, we omit the set of agents (as we assume it is always $N$, throughout all periods of time) and the endowments in each past period (as they can be obtained from the allocations, with $\sum_{i \in N} x_i^{(t)} = E^{(t)}$, for each $t\in T$). Thus, a history is simply described by the claims vectors and allocations. That is, $h=\left\{ \left( c^{(t)},x^{(t)} \right) \right\}_{t \in T}$, where $\sum_{i \in N} x_i^{(t)} \leq \sum_{i \in N} c_i^{(t)}$ for each $t \in T$. We denote by $\mathcal{H}^N$ the collection of all possible histories with agents $N$, and $\mathcal{H} \equiv \bigcup_{N \in \mathcal{N}} \mathcal{H}^N$. A \textbf{claims problem with history} (or historical claims problem) is therefore described by a 4-tuple $(N,c,E,h)$, where $(N,c,E)$ refers to the present claims problem to solve, and $h$ to the historical context (of past problems and allocations). Let $\mathcal{E}^{N}$ be the set of historical claims problems involving the set of agents $N$, and $\mathcal{E} \equiv \bigcup_{N \in \mathcal{N}} \mathcal{E}^{N}$.

For each claims problem with history $(N,c,E,h) \in \mathcal{E}$, an \textbf{allocation} is a vector $x \in \mathbb{R}^{N}$ satisfying boundedness ($0 \le x_i \le c_i$ for each $i \in N$) and balance ($\sum_{i \in N} x_i = E$). A \textbf{(general) rule} on $\mathcal{E}$ is a mapping $S: \mathcal{E} \longrightarrow \bigcup_{N \in \mathcal{N}} \mathbb{R}^{N}$ that associates with each historical claims problem $(N,c,E,h) \in \mathcal{E}$ an allocation $x = S(N,c,E,h)$. Let $\mathcal{S}$ denote the set of all (general) rules so constructed. 
Given the richness and variety of (standard) rules that have been analyzed in the literature on claims problems, it is natural to wonder whether there exist mechanisms to extend these rules from the classical framework to the class of historical claims problems. We shall refer to those mechanisms (that transform standard rules into general rules) as (extension) \emph{operators}. More specifically, an (extension) \textbf{operator} is a mapping $\Lambda: \mathcal{R} \longrightarrow \mathcal{S}$ that assigns to each standard rule $R\in\mathcal{R}$ a general rule $S=\Lambda(R)\in\mathcal{S}$.

\section{The direct approach}

\subsection{A historical operator}

For each claims problem with history $(N,c,E,h) \in \mathcal{E}$, and each $i \in N$, we denote its aggregate historical claim by $\Delta^c_i=\sum_{t=1}^{|T|} c^{(t)}_i$ and its aggregate historical allocation by $\Delta^x_i=\sum_{t=1}^{|T|} x^{(t)}_i$. Additionally, let $\widetilde{c}_i = c_i + \sum_{t=1}^{|T|}(c^{(t)}_i - x^{(t)}_i) = c_i + (\Delta^c_i - \Delta^x_i)$ be $i$'s \textbf{history-adjusted claim}. That is, $\widetilde{c}_i$ incorporates all past losses or deviations agent $i$ may have faced into the current claim. Notice that $\widetilde{c}_i \ge c_i$, as each $x^{(t)}$ is a (bounded) allocation at period $t$. Let $\Delta^c \equiv \left( \Delta^c_i \right)_{i \in N}$, $\Delta^x \equiv \left( \Delta^x_i \right)_{i \in N}$, and $\widetilde{c} \equiv \left( \widetilde{c}_i \right)_{i \in N}$ be the profiles of aggregate historical claims, allocations, and history-adjusted claims, respectively. Finally, let $\Delta \equiv \Delta^c - \Delta^x$.

In order to introduce the following (extension) operator, consider first the history-adjusted claims vector $\widetilde{c}$ as the profile of legitimate demands.\footnote{Legitimacy stands from incorporating comprehensive information about current claims as well as potential grievances from past periods.} Second, obtain a tentative allocation by applying the standard rule to the problem $(N,\widetilde{c},E) \in \mathcal{D}$. And third, make the necessary adjustments to guarantee that the final allocation satisfies balance and boundedness.\footnote{Recall that $\widetilde{c} \ge c$, so it may happen that $R_i(N,\widetilde{c},E) > c_i$ for some $i\ \in N$.}  We refer to the resulting operator so constructed as the \emph{historical operator}. Formally, 

\textbf{Historical operator}. For each $(N,c,E,h) \in \mathcal{E}$ and each $i \in N$,
\begin{equation}\label{HO} 
\Phi_i(R)(N,c,E,h) = \min\left\{ c_i, R_i(N,\widetilde{c},E) + \lambda \right\},
\end{equation}
where $\lambda \in \mathbb{R}_+$ is such that $\sum_{i \in N}  \min\left\{ c_i, R_i(N,\widetilde{c},E) + \lambda \right\} = E$.

We shall refer to $\Phi(R)$ as the general rule induced by $R$ via the historical operator $\Phi$, or, simply, the historically induced rule by $R$. The overall procedure this operator formalizes (to obtain an allocation in $\mathcal{E}$ from the application of a rule in $\mathcal{D}$) can easily be depicted:

\begin{center}
\begin{tikzpicture}
\node (a) at (0,0) {$(N,c,E,h) \in \mathcal{E}$};
\node (b) at (3.5,0) {$(N,\widetilde{c},E) \in \mathcal{D}$};
\node (c) at (7.5,0) {$R(N,\widetilde{c},E)\in\mathbb{R}^N$};
\node (d) at (12,0) {$\Phi(R)(N,c,E,h)\in\mathbb{R}^N$};

\draw[->] (a) -- (b);
\draw[->] (b) -- (c);
\draw[->] (c) -- (d);

\draw (1.75,0) node[above] {$\Delta$};
\draw (5.3,0) node[above] {$R$};
\draw (9.5,0) node[above] {$\Phi$};
\end{tikzpicture}
\end{center}

It follows from the next proposition that the historical operator is well defined. 

\begin{proposition}\label{prop_defh}
For each standard rule $R\in \mathcal{R}$, and each problem $(N,c,E,h) \in \mathcal{E}$, 
there exists $\overline{\lambda} \in \mathbb{R}_+$ 
    such that $\sum_{i \in N}  \min\left\{ c_i, R_i(N,\widetilde{c},E) + \overline{\lambda} \right\} = E$.
If $C>E$, 
$\overline{\lambda}$ is unique, whereas if $C=E$, 
$\overline{\lambda}$ is not unique, but the allocation is.
\begin{proof}
Let $R\in\mathcal{R}$ and $(N,c,E,h) \in \mathcal{E}$. We define the function $f: \mathbb{R} \longrightarrow \mathbb{R}$ as
$$
f(\lambda)=\sum_{i \in N}  \min\left\{ c_i, R_i(N,\widetilde{c},E) + \lambda \right\},
$$
for each $\lambda\in \mathbb{R}$. Clearly, $f$ is a continuous and increasing function. 
Besides, 
$$
f(0) = \sum_{i \in N}  \min\left\{ c_i, R_i(N,\widetilde{c},E) \right\} \le \sum_{i \in N}  R_i(N,\widetilde{c},E) = E.
$$
and
$$
f(C) = \sum_{i \in N}  \min\left\{ c_i, R_i(N,\widetilde{c},E) + C \right\} = \sum_{i \in N} c_i \ge E.
$$
Therefore, there exists $\overline{\lambda} \in [0,C]$ such that $f(\overline{\lambda})=E$. Now, suppose that $C \ge E$ and there exist $\overline{\lambda}_1 >  \overline{\lambda}_2$ such that $f(\overline{\lambda}_1)=f(\overline{\lambda}_2)=E$. Then, for each $i \in N$, let 
$$
x_i = \min\left\{ c_i, R_i(N,\widetilde{c},E) + \overline{\lambda}_1 \right\}
\quad \text{ and } \quad
y_i = \min\left\{ c_i, R_i(N,\widetilde{c},E) + \overline{\lambda}_2 \right\}.
$$
By construction, $\sum_{i \in N}x_i = \sum_{i \in N} y_i =E$. Now, as $\overline{\lambda}_1 > \overline{\lambda}_2$, it follows that $x_i \ge y_i \ge 0$ for each $i \in N$. Thus, $x_i=y_i$ for each $i \in N$. In addition, if $C>E$, there exists $j \in N$ such that $x_j = y_j < c_j$. Or, equivalently, $ R_j(N,\widetilde{c},E) + \overline{\lambda}_1 = R_j(N,\widetilde{c},E) +\overline{\lambda}_2$, and thus $\overline{\lambda}_1 = \overline{\lambda}_2$.
\end{proof}
\end{proposition}

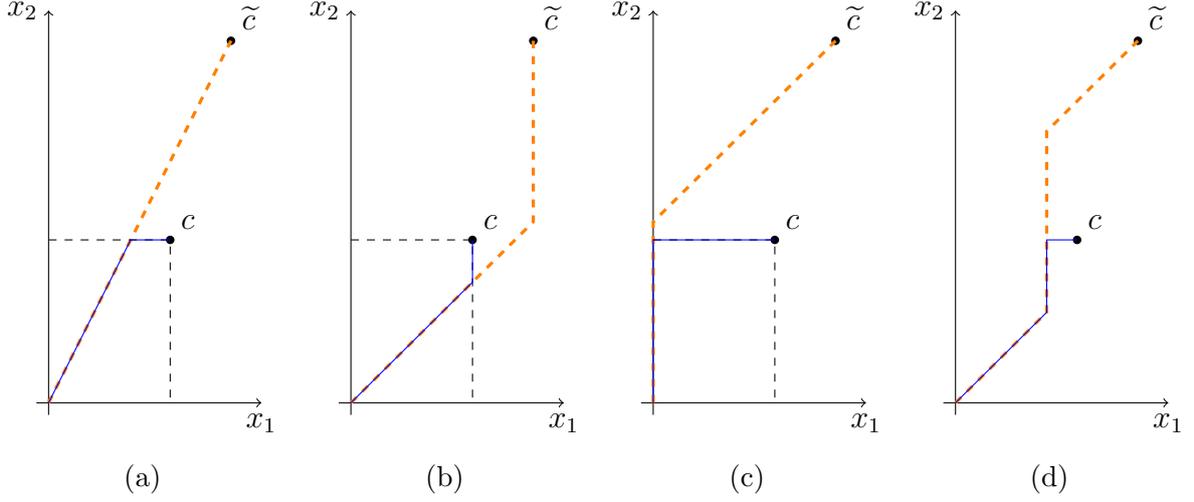
\begin{figure}
\centering
\begin{subfigure}{0.24\textwidth}
\begin{tikzpicture}[scale=0.8]
        \draw[->] (-0.2,0) -- (3.5,0) node[below] {$x_1$};
        \draw[->] (0,-0.2) -- (0,6.5) node[left] {$x_2$};
        \fill (3,6) circle (2pt) node[above right] {$\widetilde{c}$};
        \draw[very thick, orange, dashed] (0,0) -- (3,6);
        \fill (2,2.7) circle (2pt) node[above right] {$c$};
        \draw[dashed] (0,2.7) -- (2,2.7) -- (2,0);
        \draw[blue] (0,0) -- (1.35,2.7) -- (2,2.7);
\end{tikzpicture}
\caption{ }
\end{subfigure}
\begin{subfigure}{0.24\textwidth}
\begin{tikzpicture}[scale=0.8]
        \draw[->] (-0.2,0) -- (3.5,0) node[below] {$x_1$};
        \draw[->] (0,-0.2) -- (0,6.5) node[left] {$x_2$};
        \fill (3,6) circle (2pt) node[above right] {$\widetilde{c}$};
        \draw[very thick, orange, dashed] (0,0) -- (3,3) -- (3,6);
        \fill (2,2.7) circle (2pt) node[above right] {$c$};
        \draw[dashed] (0,2.7) -- (2,2.7) -- (2,0);
        \draw[blue] (0,0) -- (2,2) -- (2,2.7);
\end{tikzpicture}
\caption{}
\end{subfigure}
\begin{subfigure}{0.24\textwidth}
\begin{tikzpicture}[scale=0.8]
        \draw[->] (-0.2,0) -- (3.5,0) node[below] {$x_1$};
        \draw[->] (0,-0.2) -- (0,6.5) node[left] {$x_2$};
        \fill (3,6) circle (2pt) node[above right] {$\widetilde{c}$};
        \draw[very thick, orange, dashed] (0,0) -- (0,3) -- (3,6);
        \fill (2,2.7) circle (2pt) node[above right] {$c$};
        \draw[dashed] (0,2.7) -- (2,2.7) -- (2,0);
        \draw[blue] (0,0) -- (0,2.7) -- (2,2.7);
\end{tikzpicture}
\caption{}
\end{subfigure}
\begin{subfigure}{0.24\textwidth}
\begin{tikzpicture}[scale=0.8]
        \draw[->] (-0.2,0) -- (3.5,0) node[below] {$x_1$};
        \draw[->] (0,-0.2) -- (0,6.5) node[left] {$x_2$};
        \fill (3,6) circle (2pt) node[above right] {$\widetilde{c}$};
        \draw[very thick, orange, dashed] (0,0) -- (1.5,1.5) -- (1.5,4.5) -- (3,6);
        \fill (2,2.7) circle (2pt) node[above right] {$c$};
        \draw[blue] (0,0) -- (1.5,1.5) -- (1.5,2.7) -- (2,2.7);
\end{tikzpicture}
\caption{}
\end{subfigure}
\caption{Paths of awards for claims vector $c$ that corresponds to each of the extended rules induced by the (a) proportional, (b) constrained equal awards, (c) constrained equal losses, and (d) Talmud rules. \label{fig_paths_history}}
\end{figure}

Figure \ref{fig_paths_history} illustrates how the historical operator works for the four classical rules introduced in the previous section. In each case, the dashed orange line represents the path of awards of the standard rule $R$, from the origin to the history-adjusted claims $\widetilde{c}$. The solid blue line represents the path of awards of the induced rule $\Phi(R)$ for the present claims $c$. As we observe, in the two-agent case, $p^{\Phi(R)}(c)$ mimics $p^{R}(c)$ until the point at which one of the two agents is fully satiated, and then continues straight until reaching $c$.

\subsection{Preservation of axioms}

As mentioned above, the historical operator associates with each standard rule a new rule to solve claims problems in historical contexts. A natural question is whether the rule so constructed inherits the properties of the original rule. We say that an axiom is \emph{preserved} by the historical operator if whenever a rule $R$ satisfies it, then the historically induced rule $\Phi(R)$ satisfies the corresponding \emph{general} version of the same axiom. 

For ease of exposition, we omit the straightforward definitions of the general versions of \emph{anonymity}, \emph{continuity}, \emph{scale invariance}, \emph{self-duality}, \emph{composition up}, \emph{composition down}, \emph{securement}, \emph{resource monotonicity}, and \emph{population monotonicity}. The formulation of the remaining axioms is as follows.

\textbf{Equal treatment of equals}. For each $(N,c,E,h) \in \mathcal{E}$ and each pair $i,j\in N$, if $c_i = c_j$, $\Delta^c_i = \Delta^c_j$ and $\Delta^x_i = \Delta^x_j$, then $S_i (N,c,E,h) = S_j (N,c,E,h)$.

\textbf{Order preservation in gains}. For each $(N,c,E,h) \in \mathcal{E}$ and each pair $i,j\in N$, if $c_i \ge c_j$, $\Delta^c_i \ge \Delta^c_j$ and $\Delta^x_i \le \Delta^x_j$, then $S_i (N,c,E,h) \geq S_j (N,c,E,h)$.

\textbf{Order preservation in losses}. For each $(N,c,E,h) \in \mathcal{E}$ and each pair $i,j\in N$, if $c_i \ge c_j$, $\Delta^c_i \ge \Delta^c_j$ and $\Delta^x_i \le \Delta^x_j$, then $c_i - S_i (N,c,E,h) \geq c_j - S_j (N,c,E,h)$.



\textbf{Claims monotonicity}.  For each $(N,c,E,h) \in \mathcal{E}$ and each $i\in N$, if $c'_i > c_i$, then $S_i (N,(c'_i,c_{-i}),E,h) \geq S_i (N,c,E,h)$.

\textbf{Consistency}.  For each $(N,c,E,h)\in\mathcal{E}$, each $M\subset N,$ and each $i\in M,$ we have $S_{i} (N,c,E,h)= S_{i}(M,c_{M},E_{M},h_M),$ where $E_{M}=\sum_{i\in M} S_{i}(N,c,E,h)$.

Our first result provides the list of axioms that are preserved by the historical operator.

\begin{theorem}\label{thm_preserve} The historical operator preserves the axioms of anonymity, continuity, order preservation in gains, order preservation in losses, equal treatment of equals, scale invariance, resource monotonicity, and securement.
\begin{proof}
The preservation of \emph{anonymity} and \emph{continuity} is straightforward and, thus, we omit the proofs. We also omit the proof for \emph{equal treatment of equals} as it is a consequence of the proof for \emph{order preservation (in gains and in losses)}. We then focus on the remaining axioms.
\begin{itemize} 
    \item \emph{Order preservation (in gains and in losses)}. Let $R\in\mathcal{R}$ be a standard rule that satisfies \textit{order preservation}. Let $(N,c,E,h) \in \mathcal{E}$ and $i,j\in N$ be such that $c_i \ge c_j$, $\Delta^c_i \ge \Delta^c_j$ and $\Delta^x_i \le \Delta^x_j$. As $\widetilde{c_i}=c_i + (\Delta^c_i - \Delta^x_i) \ge c_j + (\Delta^c_j - \Delta^x_j) = \widetilde{c}_j$ and $R$ satisfies \emph{order preservation in gains}, we obtain that 
    \begin{align*}
    \Phi_i(R)(N,c,E,h) &=  \min \left\{ c_i, R_i(N,\widetilde{c},E) + \lambda \right\} \\ 
    &\ge \min \left\{ c_j, R_j(N,\widetilde{c},E) + \lambda \right\} \\
    &= \Phi_j(R)(N,c,E,h).
    \end{align*}  
    Analogously, as $R$ also satisfies \emph{order preservation in losses},
    \begin{align*}
    c_i - \Phi_i(R)(N,c,E,h) &=  c_i - \min \left\{ c_i, R_i(N,\widetilde{c},E) + \lambda \right\} = \min \left\{0, c_i - R_i(N,\widetilde{c},E) + \lambda \right\}\\
    &\ge \min \left\{0, c_j - R_j(N,\widetilde{c},E) + \lambda \right\} = c_j - \min \left\{ c_j, R_j(N,\widetilde{c},E) + \lambda \right\}\\
    &= c_j - \Phi_j(R)(N,c,E,h).
    \end{align*}
    
    
    \item \emph{Scale invariance}. Let $R\in\mathcal{R}$ be a standard rule that satisfies \emph{scale invariance}. Let $(N,c,E,h) \in \mathcal{E}$ and $\rho \in \mathbb{R}_{++}$. On the one hand, 
    $$
    \Phi(R)(N,\rho c, \rho E, \rho h) = \min \{\rho c, R(N,\rho \widetilde{c}, \rho E) + \lambda \mathbf{1}_n\},
    $$
    with $\lambda \in \mathbb{R}$ such that $\sum_{i \in N}  \min\left\{ \rho c_i, R_i(N,\rho \widetilde{c}, \rho E) + \lambda \right\} = \rho E$. On the other hand,
    $$
    \rho \Phi(R)(N,c,E,h) = \rho \min \{c,  R(N,\widetilde{c},E) + \mu \mathbf{1}_n\} = \min \{\rho c, \rho R(N,\widetilde{c},E) + \rho \mu \mathbf{1}_n\},
    $$
    with $\mu \in \mathbb{R}$ such that $\sum_{i \in N}  \min\left\{ c_i, R_i(N, \widetilde{c},E) + \mu \right\} = E$. As $R$ satisfies \emph{scale invariance}, we have that $R_i(N,\rho \widetilde{c}, \rho E) = \rho R_i(N,\widetilde{c},E)$ for each $i \in N$. Notice that 
    $$
    \min \{\rho c, R(N,\rho \widetilde{c}, \rho E) + \lambda \mathbf{1}_n\} = \min \{\rho c, \rho R(N, \widetilde{c},  E) + \lambda \mathbf{1}_n\} = \rho \min \{c, R(N,\widetilde{c}, E) + \frac{\lambda}{\rho} \mathbf{1}_n\}.
    $$
    Then,
    $$
    \sum_{i \in N}  \min\left\{ \rho c_i, R_i(N,\rho \widetilde{c}, \rho E) + \lambda \right\} = \rho E.$$
    Using \emph{scale invariance} of $R$, this is equivalent to
    $$ \sum_{i \in N}  \rho\min\left\{ c_i, R_i(N, \widetilde{c}, E) + \frac{\lambda}{\rho} \right\} = \rho E,
    $$
    and therefore
    $$ \sum_{i \in N}  \min\left\{ c_i, R_i(N, \widetilde{c}, E) + \frac{\lambda}{\rho} \right\} = E.
    $$
    By Proposition \ref{prop_defh}, we have that $\min\left\{ c_i, R_i(N, \widetilde{c}, E) + \frac{\lambda}{\rho} \right\} = \min\left\{ c_i, R_i(N, \widetilde{c}, E) + \mu \right\}$ for each $i \in N$. Therefore, $\min\left\{ \rho c_i, R_i(N, \rho \widetilde{c}, \rho E) + \lambda \right\} = \rho \min\left\{ c_i, R_i(N, \widetilde{c}, E) +\mu \right\}$, and thus $\Phi(R)(N,\rho c, \rho E, \rho h) = \rho \Phi(R)(N,c,E,h)$, as desired. 
    
    \item \emph{Resource monotonicity}. Let $R\in\mathcal{R}$ be a standard rule that satisfies \emph{resource monotonicity}. Let $(N,c,E,h) \in \mathcal{E}$ and $E'$ be such that $\sum_{i\in N}c_i\geq E'>E$. 

    Note that, alternatively to Equation (\ref{HO}), we can construct the historically induced rule $\Phi(R)$ by applying the following iterative procedure, which is based on sequentially distributing the endowment according to the standard rule $R$ and determining which claimants are satisfied. Formally, at the first stage, we define 
    $$M_{1}=\{i\in N: R_i(N, \widetilde{c}, E)> c_i\} \text{ and } \lambda_1=\sum_{i\in M_1}\left(R_i(N, \widetilde{c}, E)-c_i\right).$$ 
    At the second stage, we define 
    $$M_{2}=\{i\in N\backslash M_1:  R_i(N, \widetilde{c}, E)+\frac{\lambda_1}{| N\backslash M_1|}> c_i\}\text{ and }\lambda_2=\sum_{i\in M_2}\left(R_i(N, \widetilde{c}, E)+\frac{\lambda_1}{| N\backslash M_1|}-c_i\right).$$ 
    At the third stage, we define 
    $$M_{3}=\{i\in N\backslash \left(M_1\cup M_2\right):  R_i(N, \widetilde{c}, E)+\frac{\lambda_1}{| N\backslash M_1|}+\frac{\lambda_2}{| N\backslash \left(M_1\cup M_2\right)|}> c_i\}\text{ and }$$
    $$\lambda_3=\sum_{i\in M_3}\left(R_i(N, \widetilde{c}, E)+\frac{\lambda_1}{| N\backslash M_1|}+\frac{\lambda_2}{| N\backslash \left(M_1\cup M_2\right)|}-c_i\right).$$ We repeat this process $k$ times until $M_{k+1}=\varnothing$. Finally, let $M=M_1\cup \cdots\cup M_k$. Then, 
    $$
    \Phi_i(R)(N,c,E,h)=
    \begin{cases}
    c_i & \text{if } i\in M \\
    R_i(N, \widetilde{c}, E)+\frac{\lambda_1}{| N\backslash M_1|}+\cdots+\frac{\lambda_k}{| N\backslash \left(M_1\cup\dots \cup M_k\right)|}\footnotemark  & \text{if } i \in N\backslash M.
    \end{cases}
    $$
\footnotetext{Note that by \emph{balance} $\frac{\lambda_1}{| N\backslash M_1|}+\cdots+\frac{\lambda_k}{| N\backslash \left(M_1\cup\dots \cup M_k\right)|}=\frac{\sum_{i\in M}\left(R_i(N, \widetilde{c}, E)-c_i\right)}{|N\backslash M|}$.}
    
     As $R$ satisfies \emph{resource monotonicity}, we have, for each $i\in M_1$,
     $$R_i(N, \widetilde{c}, E')\geq  R_i(N, \widetilde{c}, E)> c_i$$
     then $ \Phi_i(R)(N,c,E',h)=c_i$ for all $i\in M_1$. By another way,  
     $$\sum_{i\in M_1}\left(R_i(N, \widetilde{c}, E')-c_i\right)\geq \sum_{i\in M_1}\left(R_i(N, \widetilde{c}, E)-c_i\right)>0$$
   therefore, it also happens that $\Phi_i(R)(N,c,E',h)=c_i$ for all $i\in M_2$. Now,
   $$\sum_{i\in M_1\cup M_2}\left(R_i(N, \widetilde{c}, E')-c_i\right)\geq \sum_{i\in M_1\cup M_2}\left(R_i(N, \widetilde{c}, E)-c_i\right)>0$$
   We can repeat this process up to step $k$ and we have $\Phi_i(R)(N,c,E',h)=c_i$  for all $i\in M$, and,
      $$\sum_{i\in M}\left(R_i(N, \widetilde{c}, E')-c_i\right)\geq \sum_{i\in M}\left(R_i(N, \widetilde{c}, E)-c_i\right)>0.$$
   Now, for any $i\in N\backslash M$, since
 \begin{align*}
    R_i(N, \widetilde{c}, E')+\frac{\sum_{i\in M}\left(R_i(N, \widetilde{c}, E')-c_i\right)}{|N\backslash M|}\ge
   R_i(N, \widetilde{c}, E)+\frac{\sum_{i\in M}\left(R_i(N, \widetilde{c}, E)-c_i\right)}{|N\backslash M|}
\end{align*}
     we can conclude that $\Phi_i(R)(N,c,E',h)\geq \Phi_i(R)(N,c,E,h)$ for all $i\in N\backslash M$.

    \item \textit{Securement}. Let $R\in\mathcal{R}$ be a standard rule that satisfies \emph{securement}. Let $(N,c,E,h) \in \mathcal{E}$. Then,
        \begin{align*}
            \Phi_i(R)(N,c,E,h)=\min\left\{  c_i, R_i(N,\widetilde{c},  E) + \lambda \right\}\geq\min\left\{  c_i, \frac{1}{|N|}\min\{\widetilde{c}_i,E\} + \lambda \right\}.
        \end{align*}
    We distinguish two cases:
    \begin{itemize}
        \item[(a)] If $\min\left\{  c_i, \frac{1}{|N|}\min\{\widetilde{c}_i,E\} + \lambda \right\}= \frac{1}{|N|}\min\{\widetilde{c}_i,E\} + \lambda$, since $\frac{1}{|N|}\min\{\widetilde{c}_i,E\} + \lambda \geq \frac{1}{|N|}\min\{c_i,E\} + \lambda \geq \frac{1}{|N|}\min\{c_i,E\}$, then $\Phi_i(R)(N,c,E,h)  \ge \min\left\{  c_i, \frac{1}{|N|}\min\{\widetilde{c}_i,E\} + \lambda \right\} = \frac{1}{|N|}\min\{\widetilde{c}_i,E\} + \lambda \geq \frac{1}{|N|}\min\{c_i,E\}$.
        \item[(b)]  If $\min\left\{  c_i, \frac{1}{|N|}\min\{\widetilde{c}_i,E\} + \lambda \right\}=c_i$, since $c_i\geq \min\{E,c_i\}\geq \frac{1}{|N|}\min\{c_i,E\}$, then $\Phi_i(R)(N,c,E,h)  \ge \min\left\{  c_i, \frac{1}{|N|}\min\{\widetilde{c}_i,E\} + \lambda \right\} = c_i \geq  \frac{1}{|N|}\min\{c_i,E\}$.
    \end{itemize}
\end{itemize}
\end{proof}
\end{theorem}

The next result identifies the axioms that are not preserved.

\begin{theorem}\label{thm_no_preserve} The historical operator does not preserve the axioms of claims monotonicity, composition up, composition down, self-duality, population monotonicity, and consistency. 
\begin{proof} 
In each case, we propose a standard rule $R$ that satisfies the axiom while its historically induced rule $\Phi(R)$ does not. 

\begin{itemize} 
    \item \emph{Claims monotonicity}. Let $\succ$ be the natural ordering of the set of potential agents $\mathbb{N}$. That is, $1 \succ 2 \succ 3 \succ \ldots $. Consider the \emph{priority rule} $R^\succ \in\mathcal{R}$ that fully honors agents' claims in the order $\succ$, until the endowment is exhausted. We then construct the (standard) rule $R\in\mathcal{R}$ that coincides with $R^\succ$ for each $(N,c,E) \in \mathcal{D}$ with $N \neq \{1,2,3\}$. If $N = \{1,2,3\}$, then 
    $$
    R(N,c,E) =
    \begin{cases}
    R^{1 \succ 3 \succ 2}(N,c,E) & \text{if } c_1,c_2>E   \\ 
    R^{3 \succ 1 \succ 2}(N,c,E) & \text{otherwise.} \\ 
    \end{cases}
    $$
    This rule satisfies \emph{claims monotonicity}, but its historical induced rule does not. Indeed, let us consider the historical claims problem $(N,c,E,h)$, where $N=\{1,2,3\}$, $c=(4,1,2)$, $E=3$, and $h=\{c^{(1)}=(3,3,3),x^{(1)}=(2,2,1)\}$. Then, $\Delta=(1,1,2)$. Moreover,
    $$
    \Phi(R)(N,c,E,h) = \min \left\{ (4,1,2), R (N,(5,2,4), 3) + \lambda \mathbf{1}_n \right\}= \left(\frac{1}{2},\frac{1}{2},2\right).
    $$
    Now, let  $c'=(4,3,2)$. Then,
     $$
    \Phi(R)(N,c',E,h) = \min \left\{ (4,3,2), R (N,(5,4,4), 3) + \lambda \mathbf{1}_n \right\}= \left(3,0,0\right).
    $$
    Thus, $\Phi_2(R)(N,c',E,h)<\Phi_2(R)(N,c,E,h)$.
    
    \item \emph{Composition up}. The standard constrained equal-losses rule satisfies \emph{composition up}, but its historically induced rule does not. To show that, let us consider the historical claims problem $(N,c,E,h) \in \mathcal{E}$, where $N=\{1,2,3\}$, $c=(10,5,2)$, $E=15$, and $h=\{c^{(1)}=(7,7,20),x^{(1)}=(2,2,2)\}$. Then, $\Delta=(5,5,18)$. Moreover, 
    \begin{align*}
    \Phi \left(R^{CEL} \right)(N,c,E,h) &= \min \left\{ (10,5,2), R^{CEL} \left(N, (15,10,20), 15\right) + \lambda \mathbf{1}_n \right\} \\
    &= \min \left\{ (10,5,2), \left( 5,0,10 \right) + \lambda \mathbf{1}_n \right\} \\
    &= (9,4,2),
    \end{align*}
    where $\lambda=4$. Now, let $E_1=5$ and $E_2=10$. Then, 
    \begin{align*}
    \Phi \left(R^{CEL} \right) (N,c,E_1,h) &= \min \left\{ (10,5,2), R^{CEL} \left(N, (15,10,20), 5\right) + \mu \mathbf{1}_n \right\} \\
    &= \min \left\{ (10,5,2), \left( 0,0,5 \right) + \mu \mathbf{1}_n \right\} \\
    &= \left( \frac{3}{2}, \frac{3}{2}, 2 \right),
    \end{align*}
    where $\mu=\frac{3}{2}$. And
    \begin{align*}
    \Phi \left(R^{CEL} \right) \left( N,c-\Phi \left(R^{CEL} \right)(N,c,E_1,h),E_2,h \right) &= \Phi \left(R^{CEL} \right) \left( N,\left(\frac{17}{2}, \frac{7}{2}, 0 \right),10,h \right) \\
    &= \left( \frac{13}{2}, \frac{7}{2}, 0 \right).
    \end{align*}
    Thus, $$\Phi \left(R^{CEL} \right)(N,c,E,h) \neq  \Phi \left(R^{CEL} \right) (N,c,E_1,h) + \Phi \left(R^{CEL} \right) \left(N,c-\Phi \left(R^{CEL} \right) (N,c,E_1,h),E_2,h \right).$$ This shows that $\Phi \left(R^{CEL} \right)$ violates \emph{composition up}.
    
    \item \emph{Composition down}. The (standard) constrained equal-losses rule also satisfies \emph{composition down}. However, its historically induced rule does not. Indeed, let $(N,c,E,h) \in \mathcal{E}$ be such that $N=\{1,2,3\}$, $c=(10,5,2)$, $E'=15$, and $h=\{c^{(1)}=(7,7,20),x^{(1)}=(2,2,2)\}$. Then, $\Delta=(5,5,18)$. Furthermore,
    \begin{align*}
    \Phi \left(R^{CEL} \right)(N,c,E',h) &= \min \left\{ (10,5,2), R^{CEL} \left( (15,10,20), 15\right) + \lambda \mathbf{1}_n \right\} \\
    &= \min \left\{ (10,5,2), \left( 5,0,10 \right) + \lambda \mathbf{1}_n \right\} \\
    &= (9,4,2),
    \end{align*}
    with $\lambda=4$. Now, let $E=9$. Then, 
    \begin{align*}
    \Phi \left(R^{CEL} \right) \left( N,\Phi \left(R^{CEL} \right)(N,c,E',h),E,h \right) &= \Phi \left(R^{CEL} \right) \left( N,(9,4,2),9,h \right)\\
    &= \min \left\{ (9,4,2), R^{CEL} \left( (14,9,20), 9\right) + \mu \mathbf{1}_n \right\} \\
    &= \min \left\{ (9,4,2), \left( \frac{3}{2},0, \frac{15}{2} \right) + \mu \mathbf{1}_n \right\} \\
    &= \left( \frac{17}{4}, \frac{11}{4}, 2 \right),
    \end{align*}
    with $\mu=\frac{11}{4}$. Using a similar reasoning, we obtain that
    $$
    \Phi \left(R^{CEL} \right)(N,c,E,h) = \left( \frac{9}{2}, \frac{5}{2}, 2 \right).
    $$
    Therefore, $$\Phi \left(R^{CEL} \right)(N,c,E,h) \neq \Phi \left(R^{CEL} \right) \left( N,\Phi \left(R^{CEL} \right)(N,c,E',h),E,h \right),$$ from where it follows that the axiom is not preserved. 
    
    \item \emph{Consistency}. The (standard) proportional rule satisfies \textit{consistency}. However, its historically induced rule violates the axiom. Indeed, let us consider the problem $(N,c,E,h) \in \mathcal{E}$, where $N=\{1,2,3,4\}$, $c=(2,4,8,6)$, $E=9$, and $h=\{c^{(1)}=(12,7,6,4),x^{(1)}=(2,2,2,2)\}$. Then, $\Delta=(10,5,4,2)$. Moreover, 
    \begin{align*}
    \Phi \left(R^{PROP} \right)(N,c,E,h) &= \min \left\{ (2,4,8,6), R^{PROP} \left( (12,9,12,8), 9\right) + \lambda \mathbf{1}_n \right\} \\
    &= \min \left\{ (2,4,8,6), \left( \frac{108}{41}, \frac{81}{41}, \frac{108}{41}, \frac{72}{41} \right) + \lambda \mathbf{1}_n \right\} \\
    &= \left( 2, \frac{269}{123}, \frac{350}{123}, \frac{242}{123} \right),
    \end{align*}
    where $\lambda=\frac{26}{123}$. 
    
    Now, let $M=\{1,2,3\} \subset N$ and $E_M=\sum_{i \in M} \Phi_i \left(R^{PROP} \right)(N,c,E,h) = \frac{865}{123}$. Then, 
    \begin{align*}
    \Phi \left(R^{PROP} \right) \left( M,c_M, E_M,h_M \right) &= \min \left\{ (2,4,8), R^{PROP} \left( (12,9,12), \frac{865}{123} \right) + \mu \mathbf{1}_m \right\} \\
    &= \min \left\{ (2,4,8), \left( \frac{3460}{1353}, \frac{865}{451}, \frac{3460}{1353}\right) + \mu \mathbf{1}_n \right\} \\
    &= \left( 2, \frac{2972}{1353}, \frac{1279}{451} \right),
    \end{align*}
    where $\mu=\frac{377}{1353}$. Thus, $\Phi_2 \left(R^{PROP} \right)(N,c,E,h) \neq \Phi_2 \left(R^{PROP} \right) \left( M,c_M, E_M,h_M \right)$, which shows that the axiom is not satisfied.

    \item \emph{Population monotonicity}. The (standard) proportional rule  satisfies \emph{population monotonicity}. However, its induced rule does not. Indeed, let $(N,c,E,h) \in \mathcal{E}$, where $N=\{1,2\}$, $c=(2,15)$, $E=15$, and $h=\{c^{(1)}=(2,20),x^{(1)}=(1,0)\}$. Thus, $\Delta=(1,20)$. Moreover, 
    $$\Phi \left(R^{PROP} \right)(N,c,E,h)=(\frac{45}{38},\frac{525}{38}).$$
    Now, let $N'=\{1,2,3\}$, $c'=(2,15,1)$, and $h'=\{c'^{(1)}=(2,20,105),x'^{(1)}=(1,0,5)\}$, then $\Delta'=(1,20,100)$. We have that
    $$\Phi \left(R^{PROP} \right)(N',c',E,h')=(2,12,1).$$
    Thus, $\Phi_1 \left(R^{PROP} \right)(N,c,E,h) < \Phi_1 \left(R^{PROP} \right)(N',c',E,h')$, but $\Phi_2 \left(R^{PROP} \right)(N,c,E,h) > \Phi_2 \left(R^{PROP} \right)(N',c',E,h')$, from where it follows that the axiom is not satisfied.  

    \item \emph{Self-duality}. The (standard) proportional rule satisfies \emph{self-duality}, but its induced rule does not. Indeed, let $(N,c,E,h) \in \mathcal{E}$,  where $N=\{1,2\}$, $c=(2,4)$, $E=2$, and $h=\{c^{(1)}=(2,2),x^{(1)}=(1,1)\}$. Thus, $\Delta=(1,1)$. Moreover, $$\Phi \left(R^{PROP} \right)(N,c,E,h)=\left( \frac{3}{4}, \frac{5}{4} \right).$$
    However, $c-\Phi \left(R^{PROP} \right)(N,c,C-E,h)=(2,4) - \left( \frac{3}{2}, \frac{5}{2} \right) = \left( \frac{1}{2}, \frac{3}{2}\right)\neq \left( \frac{3}{4}, \frac{5}{4} \right)$, from where it follows that the axiom is not satisfied.
\end{itemize}
\end{proof}
\end{theorem}

\section{The axiomatic approach}

In this section, we take a different approach by presenting axioms for extension operators. 
We shall show that a combination of these axioms leads to a characterization of the historical operator.

Our first axiom specifies a condition under which a claimant is fully honored. More precisely, if the standard rule (applied to the history-adjusted claims) stipulates that an agent should be awarded an amount exceeding his claim, then this agent should be fully honored when considering the historical context. 
Formally, 

\textbf{Present boundedness}. For each $(N,c,E,h) \in \mathcal{E}$, and each $i \in N$, if $R_i(N,\widetilde{c},E)\geq c_i$ then $\Lambda_i(R)(N,c,E,h) = c_i$.

The second axiom adapts the well-known egalitarian principle of \emph{balanced contributions} to the historical context. If two agents are rationed (i.e., they do not receive their full claims), the marginal contribution 
the operator yields 
must be the same for both claimants. Formally, 

\textbf{Balanced treatment}. For each $(N,c,E,h) \in \mathcal{E}$ and each $\{i,j\} \subset N$ such that $\Lambda_i(R)(N,c,E,h) < c_i$ and  $\Lambda_j(R)(N,c,E,h) < c_j$, 
$$
\Lambda_i(R)(N,c,E,h) - R_i(N,\widetilde{c},E) = \Lambda_j(R)(N,c,E,h) - R_j(N,\widetilde{c},E).
$$

The third axiom establishes that if the operator fully honors an agent's claim, this does not come at the expense of harming those whose claims are not fully honored. Thus, the axiom formalizes a specific form of non-arbitrariness.


\textbf{Non-arbitrariness}. For each $(N,c,E,h) \in \mathcal{E}$, and each $i \in N$ such that $\Lambda_i(R)(N,c,E,h) = c_i$ and  $R_i(N,\widetilde{c},E) < c_i$, it follows that 
$$
\Lambda_i(R)(N,c,E,h) - R_i(N,\widetilde{c},E) \leq \Lambda_j(R)(N,c,E,h) - R_j(N,\widetilde{c},E),
$$ 
for all $j \in N$ such that $\Lambda_j(R)(N,c,E,h) < c_j$.

The following result states that the combination of the last two axioms implies the first one introduced above.

\begin{lemma}\label{lemmaPB}
    If an operator satisfies balanced treatment and non-arbitrariness, then it satisfies present boundedness.
    \begin{proof}
        By contradiction, suppose $\Lambda$ is an operator that satisfies \emph{balanced treatment} and \emph{non-arbitrariness} but not \emph{present boundedness}. Then, there exists $(N,c,E,h) \in \mathcal{E}$ and $i_0 \in N$, such that $R_{i_0}(N,\widetilde{c},E)\geq c_{i_0}$ and $\Lambda_{i_0}(R)(N,c,E,h) < c_{i_0}$. As $\Lambda$ satisfies \emph{balanced treatment}, for all $j \in N$ such that $\Lambda_j(R)(N,c,E,h) < c_j$ it follows that
            $$
           \Lambda_j(R)(N,c,E,h) - R_j(N,\widetilde{c},E) = \Lambda_{i_0}(R)(N,c,E,h) - R_{i_0}(N,\widetilde{c},E) < 0.
            $$
        As $\Lambda$ and the rule $R$ are balanced, 
        $$
        \sum_{k\in N}\left(\Lambda_k(R)(N,c,E,h) - R_k(N,\widetilde{c},E) \right)=0
        $$
        Let 
        $$N^+=\left\{k \in N:\Lambda_k(R)(N,c,E,h) - R_k(N,\widetilde{c},E)\geq 0\right\},$$ and 
        $$N^-=\left\{k \in N:\Lambda_k(R)(N,c,E,h) - R_k(N,\widetilde{c},E) < 0\right\}.$$ 
        
        As $\Lambda$ satisfies boundedness, $R_k(N,\widetilde{c},E) \leq \Lambda_k(R)(N,c,E,h) \leq c_k$, for all $k \in N^+$. If $N^+ \neq \varnothing$, we consider two cases:
        \begin{enumerate}
            \item There is $k \in N^+$ such that $R_k(N,\widetilde{c},E)<c_k$. Then, we distinguish two sub-cases:
            \begin{enumerate}
                \item If $\Lambda_k(R)(N,c,E,h) <c_k$, then by \emph{balanced treatment}, $\Lambda_k(R)(N,c,E,h) - R_k(N,\widetilde{c},E) <0$, which represents a contradiction (as $k \in N^+$).
                \item If $\Lambda_k(R)(N,c,E,h) =c_k$, then by \emph{non-arbitrariness}, 
                $$
                \Lambda_k(R)(N,c,E,h) - R_k(N,\widetilde{c},E)\leq \Lambda_{i_0}(R)(N,c,E,h) - R_{i_0}(N,\widetilde{c},E) <0.
                $$
                But this is again a contradiction (as $k \in N^+$).
            \end{enumerate}
            \item There is $k \in N^+$ such that $R_k(N,\widetilde{c},E)\geq c_k$. If $R_k(N,\widetilde{c},E) > c_k$, then (as operator $\Lambda$ satisfies boundedness) $\Lambda_k(R)(N,c,E,h) - R_k(N,\widetilde{c},E) <0$, which is is a contradiction (as $k \in N^+$). 
        \end{enumerate}

        Thus, $\Lambda_k(R)(N,c,E,h)-R_k(N,\widetilde{c},E) = 0$, for each $k \in N^+$, and, therefore,
        $$
        \sum_{k\in N}\left(\Lambda_k(R)(N,c,E,h) - R_k(N,\widetilde{c},E) \right)=
        \sum_{k\in N^-}\left(\Lambda_k(R)(N,c,E,h) - R_k(N,\widetilde{c},E) \right)<0,
        $$
        which is a contradiction with the fact that $\Lambda$ and $R$ are balanced.
    \end{proof}
\end{lemma}

We are then ready to state our characterization result. 

\begin{theorem}\label{charoperator}
An operator satisfies balanced treatment and non-arbitrariness if and only if it is the historical operator.
\begin{proof}
We start by showing that the historical operator satisfies both axioms.
\begin{itemize}
    \item \emph{Balanced treatment}. Let $R \in \mathcal{R}$ be a standard rule. Let $(N,c,E,h) \in \mathcal{E}$ and let $\{i,j\} \subset N$ be such that $\Phi_i(R)(N,c,E,h) < c_i$ and  $\Phi_j(R)(N,c,E,h) < c_j$. Then,
    \begin{align*}
        \Phi_i(R) (N,c,E,h)-\Phi_j(R) (N,c,E,h) &=
         \min\left\{ c_i, R_i(N,\widetilde{c},E) + \lambda \right\}- \min\left\{ c_j, R_j(N,\widetilde{c},E) + \lambda \right\}\\
         &= R_i(N,\widetilde{c},E) + \lambda-R_j(N,\widetilde{c},E) - \lambda \\
         &= R_i(N,\widetilde{c},E) -R_j(N,\widetilde{c},E).
    \end{align*}
Thus, $\Phi_i(R)(N,c,E,h) - R_i(N,\widetilde{c},E) = \Phi_j(R)(N,c,E,h) - R_j(N,\widetilde{c},E)$.
    \item \emph{Non-arbitrariness}. Let $R \in \mathcal{R}$ be a standard rule. Let $(N,c,E,h) \in \mathcal{E}$ and let $\{i,j\} \subset N$ be such that $\Phi_i(R)(N,c,E,h) = c_i$, $R_i(N,\widetilde{c},E)<c_i$, $\Phi_j(R)(N,c,E,h) < c_j$ and $R_j(N,\widetilde{c},E)<c_j$. Then,
    \begin{align*}
        \Phi_i(R) (N,c,E,h)-\Phi_j(R) (N,c,E,h) &=
         \min\left\{ c_i, R_i(N,\widetilde{c},E) + \lambda \right\}- \min\left\{ c_j, R_j(N,\widetilde{c},E) + \lambda \right\}\\
         &= \min\left\{ c_i, R_i(N,\widetilde{c},E) + \lambda \right\}-R_j(N,\widetilde{c},E) - \lambda \\
         &\leq R_i(N,\widetilde{c},E) + \lambda-R_j(N,\widetilde{c},E) - \lambda\\
         &= R_i(N,\widetilde{c},E) -R_j(N,\widetilde{c},E),
    \end{align*}
    Thus, $\Phi_i(R)(N,c,E,h) - R_i(N,\widetilde{c},E) \leq \Phi_j(R)(N,c,E,h) - R_j(N,\widetilde{c},E)$.
\end{itemize}

Conversely, let $\Lambda$ be an operator that satisfies \emph{balanced treatment} and \emph{non-arbitrariness}. Let $R \in \mathcal{R}$ and $(N,c,E,h) \in \mathcal{E}$. Let us define $M^{\Lambda}=\{i \in N: \Lambda_i(R)(N,c,E,h) < c_i\}$, and let $i,j \in M^{\Lambda}$. By \emph{balanced treatment}, 
$$
\Lambda_i(R) (N,c,E,h)-\Lambda_j(R) (N,c,E,h)=R_i(N,\widetilde{c},E) -R_j(N,\widetilde{c},E).
$$
Adding the previous expression across agents $i \in M^{\Lambda}$, we have that
$$
\sum_{i\in M^{\Lambda}} \Lambda_i(R) (N,c,E,h) - |M^{\Lambda}|\Lambda_j(R) (N,c,E,h)= \sum_{i\in M^{\Lambda}} R_i(N,\widetilde{c},E) - |M^{\Lambda}|R_j(N,\widetilde{c},E)
$$
By definition, $\Lambda(R)$ satisfies boundedness, which implies that $\Lambda_k(R)(N,c,E,h)=c_k$ for each $k \in N \backslash M^{\Lambda}$. Then,
\begin{multline*}
\sum_{i \in N \backslash M^{\Lambda}} \Lambda_i(R)(N,c,E,h) + \sum_{i\in M^{\Lambda}} \Lambda_i(R) (N,c,E,h) - |M^{\Lambda}|\Lambda_j(R) (N,c,E,h) = \\\sum_{i \in N \backslash M^{\Lambda}} c_i + \sum_{i\in M^{\Lambda}} R_i(N,\widetilde{c},E) - |M^{\Lambda}|R_j(N,\widetilde{c},E) 
\end{multline*}
Again, by definition of rule, $\Lambda(R)$ is balanced, i.e. $\sum_{i \in N} \Lambda_i(R)(N,c,E,h)=E$. Thus,
$$
E - |M^{\Lambda}|\Lambda_j(R) (N,c,E,h) =\sum_{i \in N \backslash M^{\Lambda}} c_i + \sum_{i\in M^{\Lambda}} R_i(N,\widetilde{c},E) - |M^{\Lambda}|R_j(N,\widetilde{c},E). 
$$

Therefore,
\begin{align*}
\Lambda_j(R) (N,c,E,h) &= R_j(N,\widetilde{c},E) + \frac{1}{|M^{\Lambda}|} \left[ E - \sum_{i\in M^{\Lambda}} R_i(N,\widetilde{c},E) - \sum_{i \in N \backslash M^{\Lambda}} c_i \right] \\
&= R_j(N,\widetilde{c},E) + \frac{1}{|M^{\Lambda}|} \left[ \sum_{i\in N\backslash M^{\Lambda}} R_i(N,\widetilde{c},E) - \sum_{i \in N \backslash M^{\Lambda}} c_i \right] \\
&= R_j(N,\widetilde{c},E) + \frac{\sum_{i \in N \backslash M^{\Lambda}} \left(R_i(N,\widetilde{c},E) - c_i\right)}{|M^{\Lambda}|}\\
&=R_j(N,\widetilde{c},E) + \mu.
\end{align*}
Now consider the historical operator and let $\lambda=\frac{\sum_{i \in N \backslash M} \left(R_i(N,\widetilde{c},E) - c_i\right)}{|M|}$, where $M=\{i \in N: \Phi_i(R)(N,c,E,h) < c_i\}$. Then, 
\begin{align*}
\sum_{k \in N} \min \left\{ c_k, R_k(N,\widetilde{c},E) + \lambda\right\} &= \sum_{k \in N\backslash M} c_k + \sum_{k \in M} \left[ R_k(N,\widetilde{c},E) + \frac{\sum_{i \in N \backslash M} \left(R_i(N,\widetilde{c},E) - c_i\right)}{|M|} \right] \\
&= \sum_{k \in N\backslash M} c_k + \sum_{k \in M} R_k(N,\widetilde{c},E) + \sum_{i \in N \backslash M} \left(R_i(N,\widetilde{c},E) - c_i\right) \\
&=E
\end{align*}
Thus, in order to conclude that $\Lambda_j(R) (N,c,E,h)=\Phi_j(R) (N,c,E,h)$, we just have to show that $\mu=\lambda$. 
To do so, we reorganize agents to obtain the following chain of inequalities (which is independent of the operators):
$$
c_{(1)}-R_{(1)}(N,\widetilde{c},E) \leq c_{(2)}-R_{(2)}(N,\widetilde{c},E) \leq \ldots \leq c_{(n)}-R_{(n)}(N,\widetilde{c},E).
$$

By Lemma \ref{lemmaPB}, operators $\Lambda$ and $\Phi$ satisfy \emph{present boundedness}. Therefore, if $R_i(N,\widetilde{c},E) \geq c_i$, then $\Lambda_i(R) (N,c,E,h) = c_i$ and $\Phi_i(R) (N,c,E,h) = c_i$.

On the other hand, if there exists $(i)$ such that $\Lambda_{(i)}(R) (N,c,E,h) = c_{(i)}$ and $R_{(i)}(N,\widetilde{c},E)< c_{(i)}$, then by \emph{non-arbitrariness}, $\Lambda_{(k)}(R) (N,c,E,h) = c_{(k)}$ for all $(k)<(i)$. The same holds for operator $\Phi$. Now, let $(i_0)$ and $(i_1)$ be such that
\begin{align*}
    (i_0)=\max\{(i):\Phi_{(i)}(R)(N,c,E,h) = c_{(i)}\}\\
    (i_1)=\max\{(i):\Lambda_{(i)}(R)(N,c,E,h) = c_{(i)}\}
\end{align*}

Obviously, if $(i_0)=(i_1)$, then $M^{\Lambda} = M$, and vice versa. Altogether, we see that showing that $\mu=\lambda$ is equivalent to showing that $(i_0)=(i_1)$. By contradiction, suppose that $(i_0)\neq(i_1)$, and henceforth $\mu \neq \lambda$. Then, we have two possibilities:
\begin{enumerate}
    \item $(i_0)>(i_1)$. We distinguish two cases:
    \begin{enumerate}
    
        \item If $\mu > \lambda$, then 
        $$R_{(i_{1}+1)}(N,\widetilde{c},E)+\mu>R_{(i_{1}+1)}(N,\widetilde{c},E)+\lambda\geq c_{(i_{1}+1)},$$ 
        where the last inequality follows from the definition of $(i_0)$. Thus, $\Lambda_{(i_{1}+1)}(R)(N,c,E,h)= c_{(i_{1}+1)}$, which contradicts the definition of $(i_1)$.
        \item If $\mu < \lambda$, then 
        \begin{align*}
            \sum_{k=1}^{i_1}c_{(k)}&+\sum_{k=i_1+1}^{i_0}\left(R_{(k)}(N,\widetilde{c},E)+\mu\right)+\sum_{k=i_0+1}^{n}\left(R_{(k)}(N,\widetilde{c},E)+\mu\right) <\\
            &<\sum_{k=1}^{i_1}c_{(k)}+\sum_{k=i_1+1}^{i_0}c_{(k)}+\sum_{k=i_0+1}^{n}\left(R_{(k)}(N,\widetilde{c},E)+\mu\right) <\\
            &<\sum_{k=1}^{i_1}c_{(k)}+\sum_{k=i_1+1}^{i_0}c_{(k)}+\sum_{k=i_0+1}^{n}\left(R_{(k)}(N,\widetilde{c},E)+\lambda\right) =\\
            &=E
        \end{align*}
        which is a contradiction with the balancedness of operator $\Lambda$.
    \end{enumerate}
    \item $(i_0)<(i_1)$. We also distinguish two cases:
    \begin{enumerate}
        \item If $\mu < \lambda$, then 
        $$c_{(i_0+1)}>R_{(i_0+1)}(N,\widetilde{c},E)+\lambda>R_{(i_0+1)}(N,\widetilde{c},E)+\mu\geq c_{(i_0+1)},$$ 
        where the last inequality follows from \emph{non-arbitrariness}. 
        
        Therefore, we obtain a contradiction.
        \item If $\mu > \lambda$, then 
         \begin{align*}
\sum_{k=1}^{i_0}c_{(k)}&+\sum_{k=i_0+1}^{i_1}c_{(k)}+\sum_{k=i_0+1}^{n}\left(R_{(k)}(N,\widetilde{c},E)+\mu\right) >\\
            &>\sum_{k=1}^{i_0}c_{(k)}+\sum_{k=i_0+1}^{i_1}\left(R_{(k)}(N,\widetilde{c},E)+\lambda\right) +\sum_{k=i_0+1}^{n}\left(R_{(k)}(N,\widetilde{c},E)+\mu\right) >\\
            &>\sum_{k=1}^{i_0}c_{(k)}+\sum_{k=i_0+1}^{i_1}\left(R_{(k)}(N,\widetilde{c},E)+\lambda\right) +\sum_{k=i_0+1}^{n}\left(R_{(k)}(N,\widetilde{c},E)+\lambda\right) =\\
            &=E
        \end{align*}
        which is a contradiction with the balancedness of operator $\Lambda$.
    \end{enumerate}
\end{enumerate}
Therefore, we obtain that $(i_0)=(i_1)$, as desired. 
\end{proof}
\end{theorem}

We conclude mentioning that the two axioms in Theorem \ref{charoperator} are independent; that is, both of them are necessary to characterize the operator. 

To see that, consider, for instance, the following operator.
$$
    \Gamma^1(R)(N,c,E,h)=
    \begin{cases}
    (c_1,E-c_1), & \text{if} \quad |N|=2 \text{ and } \max\{c_1,c_2\}<E\\
    \Phi(R)(N,c,E,h), & \text{otherwise.}  
    \end{cases}
    $$
$\Gamma^1$ 
satisfies \emph{balanced treatment} but violates \emph{non-arbitrariness}. 

Likewise, consider the following operator. 
    $$
    \Gamma^2(R)(N,c,E,h)=
    \begin{cases}
    CEL(N,c,E), & \text{if} \quad |N|=2 \text{ and } \max\{c_1,c_2\}<E\\
    \Phi(R)(N,c,E,h), & \text{otherwise.}  
    \end{cases}
    $$
$\Gamma^2$ satisfies \emph{non-arbitrariness} but violates \emph{balanced treatment}.

\section{Final remarks}

We have analyzed in this paper claims problems with history. We have introduced a natural operator to extend rules from the (static) benchmark model into rules able to solve claims problems with history. The operator is determined by a three-stage procedure. First, history-adjusted claims (adding to the present-day claim all past deviations) are constructed. Second, a standard rule is applied to the resulting standard problem with those history-adjusted claims. And third, the allocation is properly adjusted to satisfy balance and boundedness. We have studied how this operator preserves properties (from the initial rule in the standard setting to its image rule in the extended setting). We have also characterized the operator by means of two natural axioms in the extended setting of claims problems with history.

Our paper may be seen as the latest attempt in a series of efforts to extend the well-known standard setting for (static) claims problems to account for their dynamic aspects (ranging from the inclusion of baselines or references in standard claims problems to consider aggregator operators). This should help address important problems we face frequently in real life. 
We mentioned several cases in the introduction that naturally call for a dynamic setting of claims problems. Other interesting cases particularly fitting that extended dynamic framework are the allocation of public resources (collected via taxes by a central government) among the regional governments of a country with a certain degree of decentralization {\citep{Chambers2019}} or the allocation of revenues collected from selling broadcasting rights of sports leagues \citep{Bergantinos2020}. In the former case, the government is supposed to approve the budget for the upcoming fiscal year each year. In the latter case, the national league association typically sets the allocation for each season yearly. And, in both cases, present-day claims might be modified by the historical grievances agents may have from the allocations in previous periods. This provide a rationale for the historical operator we have presented in this paper. 

To conclude, we mention that our work could also be seen as providing a
dynamic rationale for some classical rules to solve (standard) claims problems (such as the proportional, constrained equal awards, constrained equal losses and Talmud rules we presented at the introduction). In that sense, our contribution would be similar to \cite{Fleurbaey2011}, who provided a dynamic rationale for the three canonical axiomatic bargaining solutions (the so-called Nash, Kalai-Smorodinsky and egalitarian solutions).

\newpage

\begin{thebibliography}{45}
\providecommand{\natexlab}[1]{#1}
\providecommand{\url}[1]{\texttt{#1}}
\providecommand{\urlprefix}{URL }

\bibitem[{Acosta et~al.(2021)Acosta, Algaba, and Sánchez-Soriano}]{Acosta2021}
Acosta, R.~K., Algaba, E., and Sánchez-Soriano, J. (2021).
\newblock Multi-issue bankruptcy problems with crossed claims.
\newblock \emph{Annals of Operations Research}, 318:749--772.

\bibitem[{Acosta-Vega et~al.(2023)Acosta-Vega, Algaba, and Sánchez-Soriano}]{AcostaVega2023}
Acosta-Vega, R.~K., Algaba, E., and Sánchez-Soriano, J. (2023).
\newblock Design of water quality policies based on proportionality in multi-issue problems with crossed claims.
\newblock \emph{European Journal of Operational Research}, 311:777--788.

\bibitem[{Annabi et~al.(2012)Annabi, Breton, and François}]{Annabi2012}
Annabi, A., Breton, M., and François, P. (2012).
\newblock Game theoretic analysis of negotiations under bankruptcy.
\newblock \emph{European Journal of Operational Research}, 221:603--613.

\bibitem[{Bergantiños and Lorenzo(2008)}]{Bergantinos2008}
Bergantiños, G. and Lorenzo, L. (2008).
\newblock The equal award principle in problems with constraints and claims.
\newblock \emph{European Journal of Operational Research}, 188:224--239.

\bibitem[{Bergantiños et~al.(2010)Bergantiños, Lorenzo, and Lorenzo-Freire}]{Bergantinos2010}
Bergantiños, G., Lorenzo, L., and Lorenzo-Freire, S. (2010).
\newblock A characterization of the proportional rule in multi-issue allocation situations.
\newblock \emph{Operations Research Letters}, 38:17--19.

\bibitem[{Bergantiños and Moreno-Ternero(2020)}]{Bergantinos2020}
Bergantiños, G. and Moreno-Ternero, J.~D. (2020).
\newblock Sharing the revenues from broadcasting sport events.
\newblock \emph{Management Science}, 66:2417--2431.

\bibitem[{Bergantiños and Moreno-Ternero(2025)}]{Bergantinos2025}
---{}---{}--- (2025).
\newblock Streaming problems as (multi-issue) claims problems.
\newblock \emph{European Journal of Operational Research}.

\bibitem[{Calleja et~al.(2005)Calleja, Borm, and Hendrickx}]{Calleja2005}
Calleja, P., Borm, P., and Hendrickx, R. (2005).
\newblock Multi-issue allocation situations.
\newblock \emph{European Journal of Operational Research}, 164:730--747.

\bibitem[{Casas-Méndez et~al.(2011)Casas-Méndez, Fragnelli, and García-Jurado}]{CasasMendez2011}
Casas-Méndez, B., Fragnelli, V., and García-Jurado, I. (2011).
\newblock Weighted bankruptcy rules and the museum pass problem.
\newblock \emph{European Journal of Operational Research}, 215:161--168.

\bibitem[{Chambers and Moreno-Ternero(2017)}]{Chambers2017}
Chambers, C.~P. and Moreno-Ternero, J.~D. (2017).
\newblock Taxation and poverty.
\newblock \emph{Social Choice and Welfare}, 48:153--175.

\bibitem[{Chambers and Moreno-Ternero(2019)}]{Chambers2019}
---{}---{}--- (2019).
\newblock \emph{The Division of Scarce Resources}, pages 263--266.
\newblock Springer International Publishing.

\bibitem[{Chun(1999)}]{Chun99}
Chun, Y. (1999).
\newblock {E}quivalence of axioms for bankruptcy problems.
\newblock \emph{International Journal of Game Theory}, 28:511--520.

\bibitem[{Esta{\~{n}} et~al.(2021)Esta{\~{n}}, Llorca, Mart{\'{\i}}nez, and S{\'{a}}nchez-Soriano}]{Estan2021}
Esta{\~{n}}, T., Llorca, N., Mart{\'{\i}}nez, R., and S{\'{a}}nchez-Soriano, J. (2021).
\newblock On the difficulty of budget allocation in claims problems with indivisible items and prices.
\newblock \emph{Group Decision and Negotiation}, 30:1133--1159.

\bibitem[{Estévez-Fernández et~al.(2021)Estévez-Fernández, Giménez-Gómez, and Solís-Baltodano}]{EstevezFernandez2021}
Estévez-Fernández, A., Giménez-Gómez, J.-M., and Solís-Baltodano, M.~J. (2021).
\newblock Sequential bankruptcy problems.
\newblock \emph{European Journal of Operational Research}, 292:388--395.

\bibitem[{Fleurbaey and Roemer(2011)}]{Fleurbaey2011}
Fleurbaey, M. and Roemer, J.~E. (2011).
\newblock Judicial precedent as a dynamic rationale for axiomatic bargaining theory: Judicial precedent as a dynamic rationale.
\newblock \emph{Theoretical Economics}, 6:289--310.

\bibitem[{Giménez-Gómez and Peris(2014)}]{GimenezGomez2014}
Giménez-Gómez, J.-M. and Peris, J.~E. (2014).
\newblock A proportional approach to claims problems with a guaranteed minimum.
\newblock \emph{European Journal of Operational Research}, 232:109--116.

\bibitem[{Gutiérrez et~al.(2018)Gutiérrez, Llorca, Sánchez-Soriano, and Mosquera}]{Gutierrez2018}
Gutiérrez, E., Llorca, N., Sánchez-Soriano, J., and Mosquera, M. (2018).
\newblock Sustainable allocation of greenhouse gas emission permits for firms with leontief technologies.
\newblock \emph{European Journal of Operational Research}, 269:5--15.

\bibitem[{Hougaard et~al.(2013{\natexlab{a}})Hougaard, Moreno-Ternero, and {\O}sterdal}]{Hougaard13}
Hougaard, J., Moreno-Ternero, J., and {\O}sterdal, L.~P. (2013{\natexlab{a}}).
\newblock Rationing in the presence of baselines.
\newblock \emph{Social Choice and Welfare}, 40:1047--1066.

\bibitem[{Hougaard et~al.(2012)Hougaard, Moreno-Ternero, and {\O}sterdal}]{Hougaard12}
Hougaard, J.~L., Moreno-Ternero, J.~D., and {\O}sterdal, L.~P. (2012).
\newblock A unifying framework for the problem of adjudicating conflicting claims.
\newblock \emph{Journal of Mathematical Economics}, 48:107--114.

\bibitem[{Hougaard et~al.(2013{\natexlab{b}})Hougaard, Moreno-Ternero, and {\O}sterdal}]{Hougaard13b}
---{}---{}--- (2013{\natexlab{b}}).
\newblock Rationing with baselines: the composition extension operator.
\newblock \emph{Annals of Operations Research}, 211:179--191.

\bibitem[{Izquierdo and Rafels(2025)}]{Izquierdo2025}
Izquierdo, J.~M. and Rafels, C. (2025).
\newblock Egalitarian-in-deviation rules relative to a reference system for resolving conflicting claims problems.
\newblock \emph{Social Choice and Welfare}.

\bibitem[{Ju et~al.(2021)Ju, Kim, Kim, and Moreno-Ternero}]{Ju21}
Ju, B.-G., Kim, M., Kim, S., and Moreno-Ternero, J.~D. (2021).
\newblock Fair international protocols for the abatement of {GHG} emissions.
\newblock \emph{Energy Economics}, 94:105091.

\bibitem[{Lahiri(1994)}]{Lahiri1994}
Lahiri, S. (1994).
\newblock A compromise solution for claims problems.
\newblock \emph{European Journal of Operational Research}, 73:39--43.

\bibitem[{Mart\'inez and Moreno-Ternero(2022)}]{Martinez22}
Mart\'inez, R. and Moreno-Ternero, J.~D. (2022).
\newblock Compensation and sacrifice in the probabilistic rationing of indivisible units.
\newblock \emph{European Journal of Operational Research}, 302:740--751.

\bibitem[{Moreno-Ternero and Roemer(2006)}]{MorenoTernero06b}
Moreno-Ternero, J. and Roemer, J. (2006).
\newblock {I}mpartiality, priority, and solidarity in the theory of justice.
\newblock \emph{Econometrica}, 74:1419--1427.

\bibitem[{Moreno-Ternero and Villar(2004)}]{MorenoTernero04}
Moreno-Ternero, J. and Villar, A. (2004).
\newblock {T}he {T}almud {R}ule and the {S}ecurement of {A}gent's {A}wards.
\newblock \emph{Mathematical Social Sciences}, 47:245--257.

\bibitem[{Moreno-Ternero and Vidal-Puga(2021)}]{MorenoTernero21}
Moreno-Ternero, J.~D. and Vidal-Puga, J. (2021).
\newblock Aggregator operators for dynamic rationing.
\newblock \emph{European Journal of Operational Research}, 288:682--691.

\bibitem[{Moulin(1987)}]{Moulin87}
Moulin, H. (1987).
\newblock {E}qual or {P}roportional {D}ivision of a {S}urplus, and {O}ther {M}ethods.
\newblock \emph{International journal of Game Theory}, 16:161--186.

\bibitem[{Moulin(2000)}]{Moulin00}
---{}---{}--- (2000).
\newblock {P}riority {R}ules and {O}ther {A}symmetric {R}ationing {M}ethods.
\newblock \emph{Econometrica}, 68:643--684.

\bibitem[{O'Neill(1982)}]{ONeill1982}
O'Neill, B. (1982).
\newblock {A} problem of rights arbitration from the {T}almud.
\newblock \emph{Mathematical Social Sciences}, 2:345--371.

\bibitem[{Pulido et~al.(2008)Pulido, Borm, Hendrickx, Llorca, and S\'anchez-Soriano}]{Pulido08}
Pulido, M., Borm, P., Hendrickx, R., Llorca, N., and S\'anchez-Soriano, J. (2008).
\newblock Compromise solutions for bankruptcy situations with references.
\newblock \emph{Annals of Operations Research}, 158:133--141.

\bibitem[{Pulido et~al.(2002)Pulido, S\'anchez-Soriano, and Llorca}]{Pulido02}
Pulido, M., S\'anchez-Soriano, J., and Llorca, N. (2002).
\newblock Game theory techniques for university management: An extended bankruptcy model.
\newblock \emph{Annals of Operations Research}, 109:129--142.

\bibitem[{Roemer(1986)}]{Roemer86}
Roemer, J.~E. (1986).
\newblock Equality of resources implies equality of welfare.
\newblock \emph{Quarterly Journal of Economics}, 101:751--784.

\bibitem[{Saavedra-Nieves and Saavedra-Nieves(2020)}]{SaavedraNieves2020}
Saavedra-Nieves, A. and Saavedra-Nieves, P. (2020).
\newblock On systems of quotas from bankruptcy perspective: the sampling estimation of the random arrival rule.
\newblock \emph{European Journal of Operational Research}, 285:655--669.

\bibitem[{Thomson(1983{\natexlab{a}})}]{Thomson83a}
Thomson, W. (1983{\natexlab{a}}).
\newblock The fair division of a fixed supply among a growing population.
\newblock \emph{Mathematics of Operations Research}, 8:319--326.

\bibitem[{Thomson(1983{\natexlab{b}})}]{Thomson83b}
---{}---{}--- (1983{\natexlab{b}}).
\newblock Problems of fair division and the egalitarian principle.
\newblock \emph{Journal of Economic Theory}, 31:211--226.

\bibitem[{Thomson(2003)}]{Thomson03}
---{}---{}--- (2003).
\newblock {A}xiomatic and game-theoretic analysis of bankruptcy and taxation problems: a survey.
\newblock \emph{Mathematical Social Sciences}, 45:249--297.

\bibitem[{Thomson(2012)}]{Thomson12b}
---{}---{}--- (2012).
\newblock On the axiomatics of resource allocation: interpreting the consistency principle.
\newblock \emph{Economics and Philosophy}, 28:385--421.

\bibitem[{Thomson(2015)}]{Thomson15}
---{}---{}--- (2015).
\newblock Axiomatic and game-theoretic analysis of bankruptcy and taxation problems: {A}n update.
\newblock \emph{Mathematical Social Sciences}, 74:41--59.

\bibitem[{Thomson(2019)}]{Thomson19}
---{}---{}--- (2019).
\newblock \emph{How to divide when there isn't enough: {F}rom {A}ristotle, the {T}almud, and {M}aimonides to the axiomatics of resource allocation}.
\newblock Cambridge University Press.

\bibitem[{Thomson and Yeh(2008)}]{Thomson08}
Thomson, W. and Yeh, C.-H. (2008).
\newblock {O}perators for the adjudication of conflicting claims.
\newblock \emph{Journal of Economic Theory}, 143:177--198.

\bibitem[{Timoner and Izquierdo(2016)}]{Timoner2016}
Timoner, P. and Izquierdo, J.~M. (2016).
\newblock Rationing problems with ex-ante conditions.
\newblock \emph{Mathematical Social Sciences}, 79:46--52.

\bibitem[{van~den Brink et~al.(2013)van~den Brink, Funaki, and van~der Laan}]{Brink2013}
van~den Brink, R., Funaki, Y., and van~der Laan, G. (2013).
\newblock Characterization of the reverse talmud bankruptcy rule by exemption and exclusion properties.
\newblock \emph{European Journal of Operational Research}, 228:413--417.

\bibitem[{van~den Brink and Moreno-Ternero(2017)}]{Brink2017}
van~den Brink, R. and Moreno-Ternero, J.~D. (2017).
\newblock The reverse tal-family of rules for bankruptcy problems.
\newblock \emph{Annals of Operations Research}, 254:449--465.

\bibitem[{Young(1988)}]{Young1988}
Young, H. (1988).
\newblock Distributive justice in taxation.
\newblock \emph{Journal of Economic Theory}, 44:321--335.

\end{thebibliography}

\end{document}